\documentclass{jetpl}
\twocolumn

\usepackage{amsmath,amssymb,epsfig,color,pstricks,bm,graphics,alltt}

\usepackage[colorlinks,plainpages=false,linkcolor=black,urlcolor=blue,citecolor=black,pdfpagemode=UseNone,pdfstartview=FitBH]{hyperref}

\definecolor{MyDarkBlue}{rgb}{0,  0.3,  0.9}
\definecolor{MyDarkBlack}{rgb}{0,  0,  0}

\begin{document}

\lat

\title{Electronic structure of new iron-based superconductors: from pnictides
to chalcogenides and other similar systems}


\rtitle{Electronic structure of new iron-based superconductors}

\sodtitle{Electronic structure of iron-based superconductors:
from pnictides to chalcogenides and other similar systems}

\author{$^a$I.\ A.\ Nekrasov\thanks{E-mail: nekrasov@iep.uran.ru},
$^{a,b}$M.\ V.\ Sadovskii\thanks{E-mail: sadovski@iep.uran.ru}}

\rauthor{I.\ A.\ Nekrasov, M.\ V.\ Sadovskii}

\sodauthor{I.\ A.\ Nekrasov, M.\ V.\ Sadovskii}

\sodauthor{I.\ A.\ Nekrasov, M.\ V.\ Sadovskii}

\address{$^a$Institute for Electrophysics, Russian Academy of Sciences, Ural Branch, 
Amundsen str. 106,  Ekaterinburg, 620016, Russia\\
$^b$Institute for Metal Physics, Russian Academy of Sciences, Ural Branch, 
S. Kovalevskaya str. 18, Ekaterinburg, 620990, Russia}

\abstract{This review discusses and compares electronic spectra of new
iron-based high-temperature superconductors (HTSC), as well as a number of
chemically similar compounds. Particular attention is payed to iron chalcogenide
K$_{1-x}$Fe$_{2-y}$Se$_2$, which is isostructural to BaFe$_2$As$_2$ (122) pnictide. 
It is shown, that Fermi surfaces of K$_{1-x}$Fe$_{2-y}$Se$_2$ are essentially
different from those for pnictides. Using LDA+DMFT and LDA$'$+DMFT calculations
we show, that electronic correlations in K$_{1-x}$Fe$_{2-y}$Se$_2$ influence
the electronic structure much more significantly, than in the most studied
122 system. We also discuss the electronic structure of several multiple-band
superconductors, chemically similar to iron-based HTSC, with relatively small
values of $T_c$, such as  SrPt$_{2}$As$_{2}$, APt$_{3}$P, 
(Sr,Ca)Pd$_2$As$_2$, and non superconducting compound BaFe$_2$Se$_3$. 
It is shown, that electronic structure of these systems is very different from
previously studies iron pnictides and chalcogenides. The value of
$T_c$ in these systems can be understood within the simple BCS model.}


\maketitle

\section{Introduction}

The discovery of new class of iron-base superconductors~\cite{kamihara_08}
attracted great attention, leading to outstanding upsurge of both experimental
and theoretical works~(cf. reviews~\cite{UFN_90,Hoso_09,MazKor}).
The known at the present moment iron-based superconductors can be divided
in two classes: pnictides and chalcogenides. Below we list the typical representatives
of these classes, currently under study
(detailed references on experimental works can be found in \cite{PvsC}):
\begin{enumerate}
\item{Doped RE1111 systems (RE=La,Ce,Pr,Nd,Sm,Tb,Dy) with superconducting transition
temperature T$_c$ of the order of 25--55 K, with chemical formula 
RE O$_{1-x}$F$_x$FeAs;}
\item{Doped A122 systems (A=Ba,Sr), such as Ba$_{1-x}$K$_x$Fe$_2$As$_2$ with
T$_c$ of the order of 38 K;}
\item{The so called 111 systems like Li$_{1-x}$FeAs with T$_c\sim$ 18 K;}
\item{(Sr,Ca,Eu)FFeAs with T$_c\sim$ 36 K;}
\item{FeSe$_x$, FeSe$_{1-x}$Te$_x$ with T$_c$ up to 14;}
\item{(K,Cs)$_{1-x}$Fe$_{2-y}$Se$_2$ systems with T$_c$ up to 31K.}
\end{enumerate}

Following the discovery of HTSC in iron arsenides the intensive search of new
systems has lead to the discovery of several new superconducting systems,
which can be considered as chemical analogs of iron pnictides and iron
chalcogenides, e.g. such as
BaNi$_2$As$_2$ \cite{Bauer08},
SrNi$_2$As$_2$ \cite{Ronning08}, SrPt$_2$As$_2$ \cite{Kudo10}, SrPtAs
\cite{Elgazzar}. 
However, these systems possess rather low superconducting transition temperature
$T_c$. Recently, a number of new platinum based systems were also discovered:
APt$_3$P (A=Sr,Ca,La) \cite{Takayama} with $T_c$ equal to 8.4K, 6.6K and 1.5K 
correspondingly. Of interest is also the BaFe$_2$Se$_3$ system (Ba123) 
\cite{Ba123_1}, which was initially considered as potential superconductor
similar to (K,Cs)Fe$_2$Se$_2$ (preliminary data has shown here superconductivity
with $T_c\sim$11K). However, later it was shown, that in Ba123 superconductivity
is not observed up to 1.8K \cite{Ba123_2}. This system was shown to be an
antiferromagnetic (AFM) ``spin ladder'' with nontrivial magnetic ordering
\cite{Ba123_1,Ba123_2}. Superconductivity was also discovered in palladium
compounds: (Sr,Ca)Pd$_2$As$_2$ with  $T_c$ 0.92Ê and 1.27Ê 
correspondingly \cite{Anand}.

This short review is based upon the cycle of our investigations of electronic
structure of all these systems within the framework of density functional
theory and modern methods to take into account strong correlations, like
LDA+DMFT \cite{LDADMFT} and LDA$^\prime$+DMFT \cite{CLDA,CLDA_long} for iron
chalcogenides \cite{kfese_dmft, kfese_dmft2}, as well as upon the results of
LDA calculations for above mentioned similar systems 
\cite{srpt2as2, Ba123, apt3p, srpd2as2}. 

These works were preceded by rather long series of our works on iron pnictides
and chalcogenides \cite{Nekr,Nekr2,Nekr3,Nekr4,Nekr_kfese}, the results of these
works were discussed in details in our previous reviews \cite{UFN_90,PvsC}

All LDA calculations of band structures presented below were performed using
the basis of linearized muffin-tin orbitals (LMTO)~\cite{LMTO}, with default
parameters.

We shall see, that all systems under discussion are typical multiple-band
superconductors, making them interesting from the point of view of HTSC
search \cite{Gork,KS09}.

\section{Electronic structure and correlations in iron chalcogenides}

\subsection{KFe$_2$Se$_2$ system}

One of the most interesting iron chalcogenide systems, studied in recent years,
are the compounds like K$_{1-x}$Fe$_{2-y}$Se$_2$~\cite{Guo10}, 
Cs$_{1-x}$Fe$_{2-y}$Se$_2$~\cite{Ba123_1} and
(Tl,K)Fe$_{1-x}$Se$_{2-y}$~\cite{Fang}. It was shown experimentally, that in
K$_{0.8}$Fe$_{1.6}$Se$_2$ nontrivial AFM ordering is observed, with pretty high
Neel temperature of the order of 550 Ê. Simultaneously, practically in the
same temperature region, there is vacancy ordering in iron sublattice~\cite{Fe_order}. 
In fact, in K$_{1-x}$Fe$_{2-y}$Se$_2$ system many different phases are observed,
so that at present it is not completely clear, which phase produces superconductivity.
In a number of papers it was proposed, that the generic phase for superconductivity
is stoichiometric composition KFe$_2$Se$_2$~\cite{PvsC,Kordyuk_12_Fe_SC,Wen_12_Fe_SC},
though different points of view were also expressed~\cite{WLi_12_KFeSe_parent_for_SC}.
Note, that attempts of the synthesis of stoichiometric crystal KFe$_2$Se$_2$ 
were up to now unsuccessful. 

Important step in the studies of this iron selenide were the measurements 
ARPES spectra in K$_{0.76}$Fe$_{1.72}$Se$_2$~\cite{KFe2Se2_ARPES}. According to
these experiments, quasiparticle dispersions are significantly renormalized
due to electron correlations. For example, the experimentally determined value
of quasiparticle mass renormqalization for Fe-3d$_{xz,yz}$ orbitals is equal to
3, while for Fe-3d$_{xy}$ orbital it is equal to 10. These facts clearly require
theoretical analysis of electron -- electron interaction effects upon properties
of this system.

For the first time, the electronic spectrum of K$_{1-x}$Fe$_2$Se$_2$ in LDA
approximation was obtained in Refs.~\cite{Nekr_kfese,Shein_kfese}.

{\em Crystal structure.}

As we noted above, K$_{1-x}$Fe$_2$Se$_2$ and Cs$_{1-x}$Fe$_2$Se$_2$ systems are
structural analogs of Ba122 (cf. Ref.~\cite{Nekr2}) with following crystal
structure parameters:
K$_{1-x}$Fe$_2$Se$_2$ -- $a$=3.9136\AA~ and $c$=14.0367\AA~
and $z_{Se}$=0.3539 \cite{Guo10};
Cs$_{1-x}$Fe$_2$Se$_2$ -- $a$=3.9601\AA~, $c$=15.2846\AA~and  $z_{Se}$=0.3439 \cite{Ba123_1}.

{\em Electronic structure.}

In Fig. 1 we compare band structures and densities of states of Ba122 
\cite{Nekr2} (left) with corresponding calculation results for
K$_{1-x}$Fe$_2$Se$_2$ (black lines) and
Cs$_{1-x}$Fe$_2$Se$_2$ (gray lines) (right) for stoichiometric compositions
with  x=0 \cite{Nekr_kfese}. Comparing K$_{1-x}$Fe$_2$Se$_2$ and 
Cs$_{1-x}$Fe$_2$Se$_2$ systems we can note, that both have practically identical
electronic structures. In contrast from  Ba122 in these systems
Fe-3d and Se-4p states are separated in energy. Also in AFe$_2$Se$_2$ Se-4p 
states are 0.7 eV lower in energy, as compared to As-4p states.

\begin{figure}
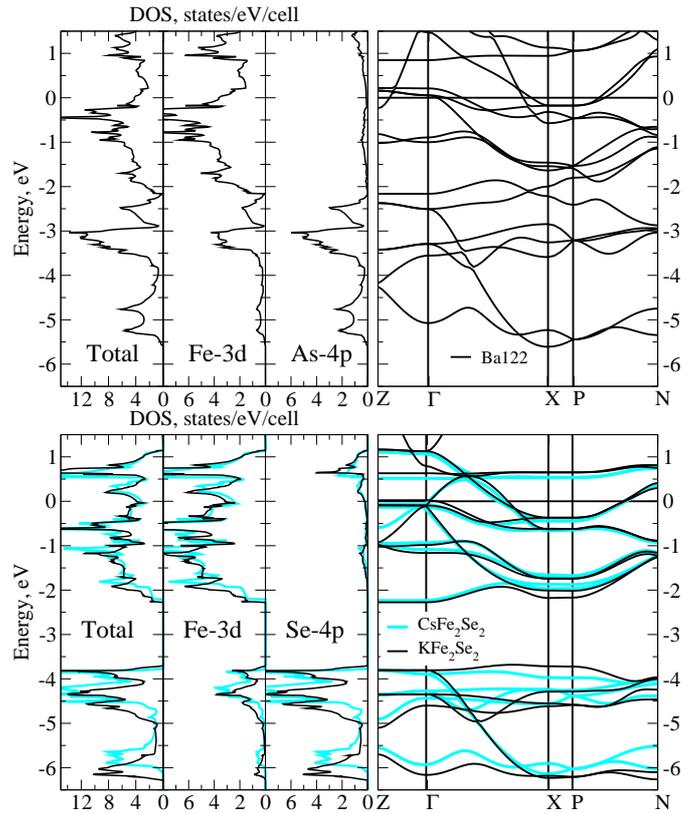

\center{
\includegraphics[clip=true,width=0.5\textwidth]{Ba122_DOS_bands.eps}
\includegraphics[clip=true,width=0.5\textwidth]{KCs_DOS_bands.eps}
}
\caption{Fig. 1. Band dispersions and densities of states for Ba122 (left),
KFe$_2$Se$_2$ (black lines) and CsFe$_2$Se$_2$ (gray lines)(right), 
obtained in LDA calculations. Fermi level is at zero energy.}
\label{bands_kfese}
\end{figure}

Similarly to Ba122 in chalcogenides under discussion practically only Fe-3d 
states cross the Fermi level. Also similarly to Ba122 \cite{Nekr2} 
and other iron pnictides, the main contribution to density of states
at the Fermi level comes from $t_{2g}$ states ($xy$, $xz$ è $yz$).
The states of $e_g$ symmetry ($3z^2-r^2$ and $x^2-y^2$) practically do not
contribute at the Fermi level. Main difference between Ba122 and new systems
is observed near the $\Gamma$-point. In Z-$\Gamma$ direction for 
(K,Cs)Fe$_2$Se$_2$ systems the anti-bonding part of Se-4p$_z$ orbital forms an
electronic pocket. In Ba122 corresponding band is 0.4 eV higher in energy an
is much steeper, so that this band passes far enough from the $\Gamma$-point. 
However, if we dope (K,Cs)Fe$_2$Se$_2$ by holes, we obtain the band structure in
the vicinity of the Fermi level, which is similar to bands in Ba122
(at 60\% hole doping) -- three hole cylinders, while the stoichiometric system
KFe$_2$Se$_2$ possesses one small electronic pocket and large, hole-like
pocket, while in CsFe$_2$Se$_2$ there is only one electronic pocket around the
$\Gamma$-point.

In Fig. 2 we show Fermi surfaces obtained from LDA calculations for
K$_{1-x}$Fe$_2$Se$_2$ (above) and Cs$_{1-x}$Fe$_2$Se$_2$ (below) for
different doping levels: x=0 -- left, x=0.2 -- in the middle, and x=0.6 -- right.
In all cases we observe two practically two-dimensional electron sheets at the
corners of the Brillouin zone, independent of the doping level.
Main difference in comparison to Ba122 pnictide is observed for
(K,Cs)Fe$_2$Se$_2$ systems around the $\Gamma$-point, especially for
x=0 and x=0.2. For doping level x=0.6 both K and Cs selenides possess Fermi
surfaces similar to those in Ba122 pnictide (cf. Fig. 2 and Ref.~\cite{Nekr2}) 
with typical hole-like cylinders at the center of Brillouin zone.

Thus, for doping levels corresponding to superconducting phase, the topology of
Fermi surfaces is quite different from that in Ba122. In particular, there is
no ``nesting'' of electronic and hole Fermi surfaces at all
(the property of ``nesting'' is considered in many works on pnictides as an important property for 
explanation of their magnetic and superconducting properties).

\begin{figure}
\center{
\includegraphics[clip=true,width=0.4\textwidth]{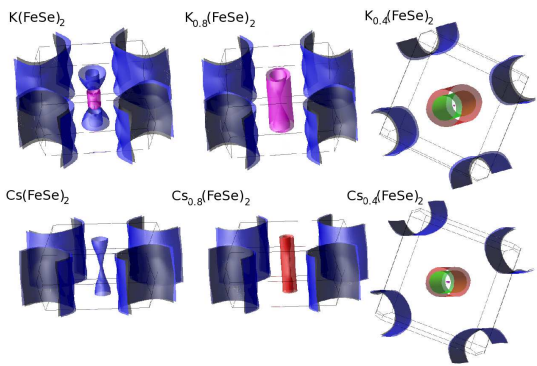}
}
\caption{Fig. 2. Fermi surfaces obtained from LDA calculations for
K$_{1-x}$Fe$_2$Se$_2$ (above) and Cs$_{1-x}$Fe$_2$Se$_2$ (below) for different
doping levels: x=0 -- left, x=0.2 -- in the middle, and x=0.6 -- right.
}
\label{fs_kfese}
\end{figure}

In Ref.~\cite{Ba123_1} it was shown that the temperature of superconducting 
transition T$_c$ of K and Cs materials under study depends of the height of Se
ion above Fe ions plane~\cite{hPn2}. This dependence was studied in details
in our work~\cite{Kucinskii10}. Let us make simple estimates of $T_c$
using the standard BCS expression for $T_c$: $T_c=1.14\omega_D e^{-2/gN(E_F)}$.
The value of the total density of states at the Fermi level N($E_F$) 
for  KFe$_2$Se$_2$ is equal to 3.94~states/eV/cell, while for  äëÿ CsFe$_2$Se$_2$ 
it is 3.6 states/eVýÂ/cell. Taking the value of Debye frequency
$\omega_D$=350K and the coupling constant $g$=0.21eV, obtained for Ba122 
(cf. Ref.~\cite{Kucinskii10}), we get T$_c$=34K and 28.6K for K and Cs 
selenides correspondingly (the ratio of $T_c$ for these systems
is equal to 1.18). These estimates agree rather well with experimental data of
31K \cite{Guo10} and 27K ($T_c$ ratio 1.15) \cite{Ba123_1}. Smaller values of
$T_c$ in CsFe$_2$Se$_2$ can be explained by isotope effect. Taking into
account that  60\% hole doping leads to higher values of N($E_F$) for both
systems, which are (according to our calculations) 4.9~states/eV/cell in
K selenide and  4.7~states/eV/cell in Cs selenide, we can expect, that 
corresponding temperatures of superconducting transition for these systems at 
this doping level will be correspondingly 57K and 52K.

Note that these estimates do not necessarily assume the electron-phonon
mechanism of superconductivity, and the value of $\omega_D$ can be referred to 
any Bosonic excitation, responsible for pairing, e.g. to spin fluctuations.

\subsection{K$_{0.76}$Fe$_{1.72}$Se$_2$ system}

To study correlation effects in K$_{0.76}$Fe$_{1.72}$Se$_2$ system \cite{kfese_dmft}, 
as well as correlation effects in K$_{1-x}$Fe$_{2-y}$Se$_2$ at different hole
dopings \cite{kfese_dmft2}, we have used the standard LDA+DMFT \cite{LDADMFT} 
approach and its modification --- LDA$'$+DMFT, developed by us \cite{CLDA,CLDA_long}, 
allowing consistent solution of the ``double counting'' problem of Coulomb
interaction \cite{dcproblem}.

First of all, let us discuss the band structures for stoichiometric composition
KFe$_2$Se$_2$, obtained within LDA and LDA$'$ calculations. Similarly to the
case of transition metal oxides (cf. Refs. \cite{CLDA,CLDA_long})
LDA$'$ calculations for KFe$_2$Se$_2$ produce energy dispersions at the Fermi
level close to those obtained within LDA. However, there are certain deviations
near $\Gamma$-point. There is a shift of Se-4p band in  LDA$'$ calculations
down in energy with respect to LDA bands of the order of  0.5 eV 
(cf. Fig. 4 and Ref. \cite{kfese_dmft2}).

For DMFT calculations we have taken the values of Coulomb and exchange 
interactions at Fe-3d orbitals to be equal to $U=3.75$~eV and $J=0.56$ eV, 
correspondingly, in agreement with Ref.~\cite{KFe2Se2_ARPES}.
inverse temperature $\beta=40$ eV$^{-1}$ (290 K), the number imaginary time
intervals $L=180$ and the number of pseudospin flips $\sim 10^6$ to reach 
self-consistency. For K$_{0.76}$Fe$_{1.72}$Se$_2$ the total number of electrons
in DMFT calculations was taken to be n$_e$=26.52.

For the given doping level the account of local Coulomb interaction has lead
to the following changes close to the Fermi level \cite{kfese_dmft}: 
practically nothing has changed for Fe-3d$_{x^{2}-y^{2}}$ orbitals,
for Fe-3d$_{3z^{2}-r^{2}}$ states the Hubbard band appeared above the Fermi 
level, while for Fe-3d$_{xz,yz}$ and Fe-3d$_{xy}$ orbitals we observe the
strong change in the density of states, reminding pseudogap behavior. 
Thus, even from the analysis of the density of states only, we can conclude,
that correlation effects in K$_{0.76}$Fe$_{1.72}$Se$_2$ compound are quite
relevant. At the same time, their role for different orbitals may be quite
different. Most important these effects are for Fe-3d$_{xz,yz}$ and 
Fe-3d$_{xy}$ orbitals.

In Fig. 3 we compare spectral functions, calculated within LDA+DMFT 
(left side, lower figure) and LDA$'$+DMFT (right side, lower figure) with angle
resolved photoemission (ARPES) spectra for different types of polarization
(upper figures)~\cite{KFe2Se2_ARPES} for chemical composition 
K$_{0.76}$Fe$_{1.72}$Se$_2$.
\begin{figure}[!ht]
\center{\includegraphics[width=0.45\textwidth]{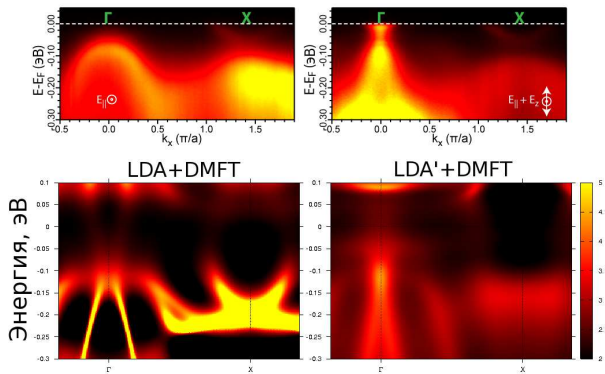}}
\caption{Fig. 3. Comparison of spectral functions obtained within 
LDA+DMFT (left side, below) and LDA$'$+DMFT (right side, below) calculations
with ARPES spectrum (above) for K$_{0.76}$Fe$_{1.72}$Se$_2$. Intensity of
spectral function is shown by color. Zero energy is at the Fermi level.}
\label{FIG:CLDA_DMFT:KFe2Se2:SF_with_ARPES}
\end{figure}
In the left upper part of Fig.3 we show the ARPES spectrum, obtained with light
polarized along the iron layers, while in the right upper part of the figure
we show the ARPES spectrum, obtained from the sum of polarizations along and
across iron layers. In the lower part of Fig. 3 we show spectral functions,
obtained within LDA+DMFT and LDA$'$+DMFT calculations for 
$\Gamma-X$-direction, studied in the experiments for the same interval of
energies.

Both spectral functions and ARPES spectra show, that close to the Fermi level
in energy interval $\pm0.05$ eV there are no well-defined maxima of intensity,
besides the vicinity of $\Gamma$-point in the experiments with full polarization.
This fact indicates, that K$_{0.76}$Fe$_{1.72}$Se$_2$ is a bad metal and
confirms the pseudogap behavior, observed in the density of states
\cite{kfese_dmft}. High intensity region near $X$-point in  ARPES spectra is
present in the energy interval $(-0.25,-0.1)$ ýÂ, where we observe
quasiparticle bands in LDA+DMFT and LDA$'$+DMFT spectral functions. 
Close to $\Gamma$-point in one of the ARPES spectra we see the wide energy
dispersion with broad maximum, while in other ARPES spectrum we observe
energy dispersion with more sharp maximum, which extends up to the Fermi level.
In calculated spectral densities in this energy interval we obtain two 
quasiparticle bands. The form of quasiparticle band with wide maximum is
almost the same for the spectral functions obtained within LDA+DMFT and 
LDA$'$+DMFT, while the form of quasiparticle band with narrow maximum in
obtained LDA$'$+DMFT calculations is closer to energy dispersion observed
in ARPES spectrumå (cf. Fig. 3).

Thus we conclude, that for iron selenide K$_{0.76}$Fe$_{1.72}$Se$_2$ 
correlation effects play an important role. They lead to significant changes if
LDA energy dispersion. In contrast to iron arsenides, where quasiparticle
bands close to the Fermi level are well-defined~\cite{Haule,Craco,Shorikov,Ba122_DMFT}, 
in K$_{0.76}$Fe$_{1.72}$Se$_2$ system we observe strong suppression of 
quasiparticle bands at the Fermi level. This confirms that correlation effects
in K$_{0.76}$Fe$_{1.72}$Se$_2$ are stronger, than in iron arsenides.
The value of quasiparticle renormalization (correlation narrowing) of bands
at the Fermi level is equal to 4-5, while in iron arsenides this factor is
equal to 2-3 \cite{Ba122_DMFT} for the same values of interaction parameters.

As electronic properties of K$_{0.76}$Fe$_{1.72}$Se$_2$ system in the vicinity
of Fermi level are strongly affected by correlation effects, the question arises:
how correlation effects change at different doping levels in
K$_{1-x}$Fe$_{2-y}$Se$_2$ compounds. To study this problem, in  
Ref.~\cite{kfese_dmft2} we have chosen three doping levels: one corresponding to
stoichiometric composition (n$_e$=29.00), and two others with total electron
numbers n$_e$=28.00 and n$_e$=27.20. As for K$_{0.76}$Fe$_{1.72}$Se$_2$ our
results obtained within LDA$'$+DMFT approach were shown to be in better
agreement wit experimental ARPES spectra, we have used the same LDA$'$+DMFT
approach in our calculations for other doping levels.
It was shown, that the form and position of the same quasiparticle bands 
is significantly transformed under doping. Position of quasiparticle bands
with respect to each other also changes. Thus, depending on the doping level
the role of correlation effects changes for each of the quasiparticle bands.
To determine the influence of local Coulomb interaction upon each of 
Fe-3d bands we have calculated correlation renormalization factors and the
energy shifts of LDA$'$ bands to match the maxima of spectral functions
obtained in LDA$'$+DMFT calculations for all four doping levels.
\begin{figure}[h]
\center{\includegraphics[width=0.45\textwidth]{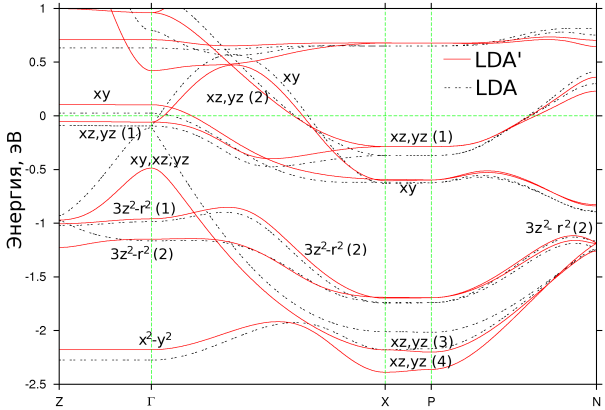}}
\caption{Fig. 4. Energy dispersions from LDA$'$ calculations (red lines) in
KFe$_2$Se$_2$ along symmetry directions in the Brillouin zone with 
notations for those bands, for which renormalization factors were fitted.
Black lines -- energy dispersions from LDA calculations. Zero energy
corresponds to the Fermi level.}
\label{FIG:CLDA_DMFT:KFe2Se2:bands_for_compression}
\end{figure}
Renormalization factors and energy shifts (in parentheses) for LDA$'$ bands are
listed in Table~1. Energy shifts are given for LDA$'$ bands in electronvolts.

Black dotted lines in Fig. 4 represent LDA energy dispersions, which both in
form and position are similar to energy dispersions obtained in LDA$'$ 
calculations. Thus, renormalization factors in LDA+DMFT will be approximately
the same as in LDA$'$+DMFT calculations (Table~1). Results for
K$_{0.76}$Fe$_{1.72}$Se$_2$ (n$_e$=26.52) composition confirm this conclusion.

Energy dispersions of LDA$'$ calculation in the interval $(-1.0,0.4)$ eV 
are stronger renormalized by iïteractions, than energy dispersions below
$-1.0$ eV. Correlation renormalization factors for 3z$^2$-r$^2$ (2), xz, yz (3), 
xz, yz (4) and x$^2$-y$^2$ bands (Fig. 4 and Table~1), which in LDA$'$ 
calculations are below $-1.0$ eV, are close to renormalization factor of
of Fe-3d band as a whole. Similarly, the xz, yz, xy, band, which belong to
energy interval $(-1.0,-0.5)$ eV, is weakly renormalized as the whole
Fe-3d band and its correlation renormalization factor does not depend on
doping. The values of correlation renormalization for
xz, yz (2) band is weakly dependent on doping and is approximately equal to
$2.5$, despite the fact, that this band is rather close to the Fermi level.
Renormalization factor for 3z$^2$-r$^2$ (1) band increases sharply up to
the value of $4.7$ for filling n$_e$=26.52.
\begin{table*}[!ht]
\caption{Table 1. The values of correlation renormalization factors for
separate LDA$'$ bands, deboted in Fig.~\ref{FIG:CLDA_DMFT:KFe2Se2:bands_for_compression}. 
In parentheses -- energy shifts on the scale of LDA$'$ bands in eV.}
\begin{center}
\begin{tabular}{|c|c|c|c|c|}
\hline
Orbital     & n$_e$=26.52 & n$_e$=27.20  & n$_e$=28.00 & n$_e$=29.00     \\
nature  &  &  &  &  \\
\hline
xy LHB          & 1.5 (-0.23) & 3.9 (-0.73)  & 2.65 (-0.61) & 1.7 (-0.35) \\
xy UHB          & no         & $\sim$4.0 (+0.60) & 1.7 (+0.25) & $\sim$4.0 (+0.75)  \\
                &             &              & part (top) & \\
\hline
xz, yz (1) LHB  & 4.2 (-0.78) & 3.0 (-0.75)  & 2.6 (-0.69) & 1.7 (-0.38)  \\
xz, yz (1) UHB  & 1.6 (+0.19) & $\sim$2.5 (+0.48) & 3.0 (+0.56) & 4.0 (+0.77)  \\
\hline
xz, yz (2)      & 2.3 (-0.48) & $\sim$2.5 (-0.60) & 2.6 (-0.69) & 1.7 (-0.38)  \\
\hline
xz, yz, xy      & 1.2 (-0.1)  & 1.3 (-0.09)  & 1.3 (-0.10) & 1.4 (-0.17)  \\
\hline
3z$^2$-r$^2$ (1)& 4.7 (-0.85) & 2.0 (-0.30)  & 1.3 (-0.03) & 1.25 (0.00)  \\
3z$^2$-r$^2$ (2)& 1.1 (+0.25) & 1.3 (0.00)   & 1.3 (-0.03) & 1.25 (0.00)  \\
\hline
xz, yz (3)      & 1.1 (+0.10) & 1.1 (+0.17)  & 1.0 (+0.40) & 1.4 (-0.10)  \\
xz, yz (4)      & 1.1 (+0.10) & 1.1 (+0.15)  & 1.0 (+0.35) & 1.4 (-0.17)  \\
\hline
x$^2$-y$^2$     & 1.1 (+0.20) & 1.0 (+0.32)  & 1.3 (-0.07) & 1.3 (-0.07)  \\
\hline
Fe3d band     & 1.3         & 1.3          & 1.25        & 1.25         \\
as a whole    &  &  &  &  \\
\hline
\end{tabular}
\end{center}
\label{table:CLDA_DMFT:KFe2Se2_compression}
\end{table*}

For xz, yz (1) band the value of correlation renormalization in the lower
Hubbard band monotonously grows with doping, while in the upper Hubbard band
it monotonously decreases. For xy band dependence on doping is nonmonotonous.
In its lower Hubbard band correlation effects grow with hole doping and reach
maximum at n$_e$=27.20, then decrease.  In upper Hubbard xy band for 
n$_e$=29.00 and n$_e$=27.20 the value of correlation renormalization
is large enough --- about $4.0$, while for n$_e$=28.00 it is smaller --
1.7, and for n$_e$=26.52 the upper Hubbard xy band is absent.

Thus, for K$_{1-x}$Fe$_{2-y}$Se$_2$ compound the hole doping from stoichiometric
composition KFe$_2$Se$_2$ up to K$_{0.76}$Fe$_{1.72}$Se$_2$ leads to the growth
of correlation effects for the same values of direct (Hubbard)
$U$ and exchange (Hund) $J$ interactions. These effects are different for
different quasiparticle bands and in different parts of the Brillouin zone. 

Possibly the main expression of more strong role of correlation effects in this
system, in contrast to previously studied iron based superconductors, is the
absence of well-defined quasiparticles bands in the vicinity of the Fermi
level.

\section{New superconductors --- chemical analogs of iron pnictides}

\subsection{SrPt$_2$As$_2$ system}

{\em Crystal structure.}

In Ref.~\cite{Imre07} it was shown, that SrPt$_2$As$_2$ has orthorhombic
structure with space symmetry group $Pmmn$. In some sense this system reminds
tetragonal crystal structure of CaBe$_2$Ge$_2$-type with space symmetry
group $P4/nmm$. We show this structure in Fig. 5. There are two layers of
PtAs$_4$ and AsPt$_4$ tetrahedra in elementary cell. In one of these layers
square lattice is formed by platinum ions, while in another by arsenic.
Corresponding tetrahedrons are shown in Fig. 5 by lines.
From the point of view of chemical formula SrPt$_2$As$_2$ is similar to HTSC
system BaFe$_2$As$_2$ \cite{rott}. However, in the elementary cell of 
this last case there are two mirror reflected layers of FeAs$_4$ tetrahedrons \cite{Nekr2}.
Note also, that  majority of iron pnictides have $P4/nmm$ structure \cite{Kucinskii10}.
In our calculations we have used an idealized tetrahedral structure with
$P4/nmm$ group (cf. details in Ref. \cite{srpt2as2}). Parameters of tetrahedral
crystal structure can be determined averaging $a$=4.482\AA~ and $b$=4.525\AA~ 
for orthorhombic phase and taking $c$=9.869\AA.

\begin{figure}
\center{
\includegraphics[clip=true,width=0.4\textwidth]{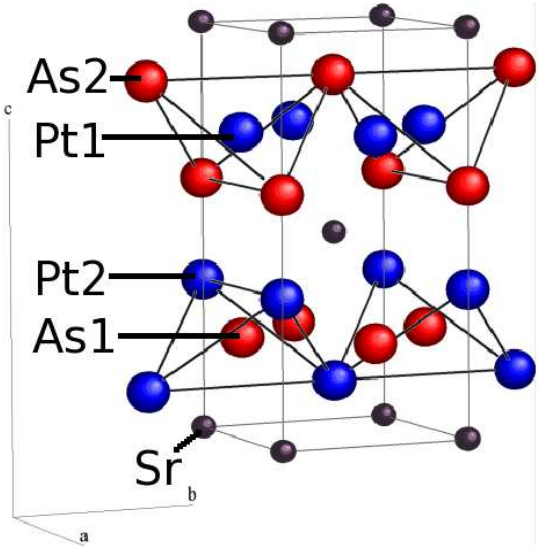}
}
\label{fig1_srptas}
\caption{Fig. 5. Idealized tetragonal crystal structure of SrPt$_2$As$_2$.}
\end{figure}

\begin{figure}
\center{
\includegraphics[clip=true,width=0.4\textwidth]{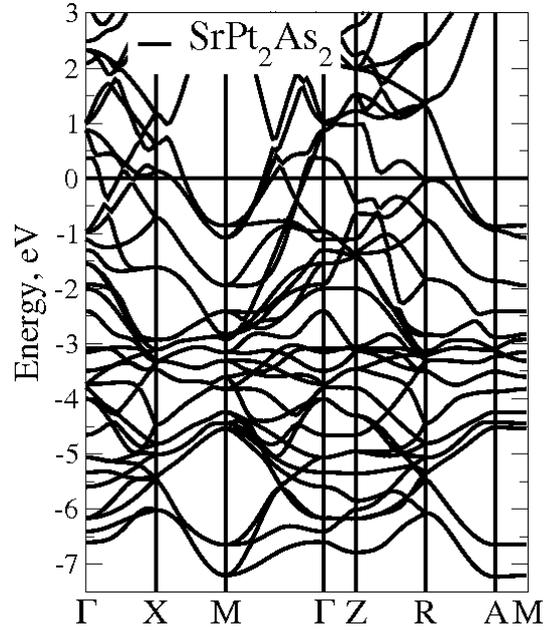}
}
\label{band_srptas}
\caption{Fig. 6. LDA electronic dispersions for SrPt$_2$As$_2$.
Zero energy at the Fermi level.}
\end{figure}

{\em Electronic structure.}

In Fig. 6  we show electronic bands calculated in LDA along directions of high
symmetry in the first Brillouin zone of SrPt$_2$As$_2$. Note that SrPt$_2$As$_2$
bands close to M-point have some similarity with 1111 FeAs system \cite{Nekr,Nekr4}.
However, close to the Fermi level bands are completely different from those in
1111 and 122 systems. As in SrPt$_2$As$_2$ there are several bands crossing
the Fermi level, we can speak about essentially multiple-band nature of
superconductivity in this system: four bands cross Fermi level in
$\Gamma$-X direction and six in M-$\Gamma$ direction.

As was shown in Ref. \cite{srpt2as2} the dominant contribution at the Fermi level
is from Pt1-5d states. However, the joint contribution of Pt2-5d and As1-4p, 
As2-4p states to the density of states at the Fermi level $E_F$ is also
significant enough. This fact makes SrPt$_2$As$_2$ different from HTSC
iron pnictidesîò ÂÒÑÏ, where As-4p states are practically absent at the Fermi
level \cite{Nekr,Nekr2,Nekr3,Nekr4}. The main part of the spectral weight of
Pt-5d states is situated well below the Fermi level, which is connected with
larger number of valence electrons in Pt, as compared to Fe. Obviously,
Pt-5d states are more extended and produce wider band as compared to Fe-3d.

In Fig. 7 we show the complex picture of the Fermi surface for this system, 
obtained in LDA approximation. The sheets of this Fermi surface are essentially
three-dimensional, which sharply differs SrPt$_2$As$_2$ from 1111 and 122 
pnictides.

Thus we see, that SrPt$_2$As$_2$ has a complex band structure close to the Fermi
level and complicated multi-sheet topology of the Fermi surface, which strongly
differs from those observed in iron pnictides. In general case, the 
multiple-band system can support very complicated types of Cooper pairing with
superconducting gaps of different sizes on different sheets of the Fermi
surface, e.g. as in FeAs compounds \cite{Gork,KS09}. From general symmetry
analysis \cite{VG,SU} it is known, that for tetragonal symmetry and singlet
Cooper pairs either isotropic or anisotropic $s$-wave pairing is possible,
as well as several types of $d$-wave pairing. However, from symmetry 
considerations only nothing can be said about $s^{\pm}$-type of pairing
with isotropic gaps of different signs on different sheets of the Fermi
surface, which is considered to be most probable for iron pnictides \cite{UFN_90,KS09}.

Let us make simplest estimates of the $T_c$ value within BCS theory.
The value of density of states at the Fermi level N($E_F$), obtained in our
calculations is equal to 5.6 states/eV/cell. Then the coefficient of 
linear  temperature contribution to specific heat following from the results
of LDA calculations will be $\gamma_b=\frac{\pi^2}{3}N(E_F)$=
13.1 mJ/mole/K$^2$, which is in reasonable agreement with experimental estimates
giving 9.7 mJ/mole/K$^2$ \cite{Kudo10}. Let us estimate the pairing coupling
constant $\lambda$ in BCS expression for $T_c=1.14\omega_D e^{-1/\lambda}$,
using the experimental value of Debye frequency $\omega_D$.
For $\omega_D$=200K and $T_c$=5.2K~\cite{Kudo10} we get for 
SrPt$_2$As$_2$ $\lambda$=0.26. Then we can estimate $T_c$ for isovalent
superconducting systems BaNi$_2$As$_2$ \cite{Bauer08} and
SrNi$_2$As$_2$ \cite{Ronning08} with $T_c$ equal to 0.7K \cite{Bauer08} 
and 0.62K \cite{Ronning08} respectively. In this case we need the values of
total LDA densities of states for BaNi and SrNi systems. Our calculations have
given N($E_F$)=3.86~states/eV/cell for BaNi and N($E_F$)=2.81~states/eV/cell 
for SrNi system, which agree well with other calculations \cite{Subedi08,Chen09}.
Changing the value of $\lambda$ proportionally to the change
of the value of N($E_F$) we get $T_c$ for these systems equal to 0.97K and 
0.13K respectively, in good agreement with experiments \cite{Bauer08,Ronning08}.

\begin{figure}
\center{
\includegraphics[clip=true,width=0.35\textwidth]{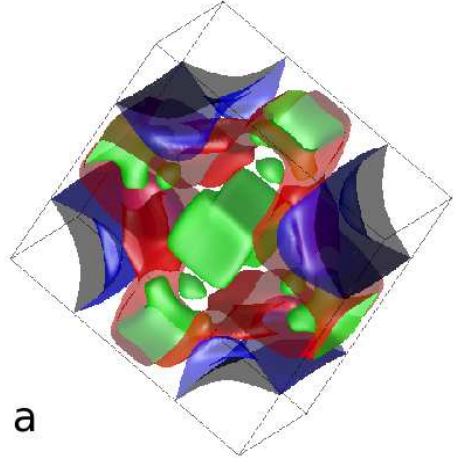}
}
\label{fs_srptas}
\caption{Ðèñ. 7. Fermi surface of SrPt$_2$As$_2$ obtained from LDA calculations.}
\end{figure}

\subsection{APt$_3$P system}

Another new platinum based system -- APt$_3$P (A=Sr,Ca,La) was discovered and
described in Ref.  \cite{Takayama}, where the values of $T_c$ 
were obtained to be  8.4K, 6.6K and 1.5K respectively.
In this section we consider the electronic structure of
SrPt$_3$P system, first obtained in Ref. \cite{apt3p}.

{\em Crystal structure.}

Crystals of SrPt$_3$P belong to tetragonal symmetry group {\it P4/nmm}
with $a$=5.8094\AA~and $c$=5.3833\AA~\cite{Takayama} (cf. Fig. 8).
Between strontium ions layers sit ``antiperovskite''  Pt$_6$P octahedra,
with Pt1 ions in layers occupy 4e (1/4,1/4,1/2) positions, while apex
Pt2 ions occupy -- 2c (0,1/2,0.1409). Phosphorous inside octahedra also
occupy 2c positions with $z$=0.7226. Note that these Pt$_6$P octahedra
are not ideal: distances from PPt$_2$ layer to different apex Pt2 ions are
different, while inlayer Pt1 ions form ideal square. In Fig. 8 it is clearly
seen, that Pt$_6$P octahedra, having the common facets,form two-dimensional
plane with square lattice. In the following we assume, that LaPt$_3$P 
has the crystal structure identical to that of SrPt$_3$P.

\begin{figure}
\center{
\includegraphics[clip=true,width=0.4\textwidth]{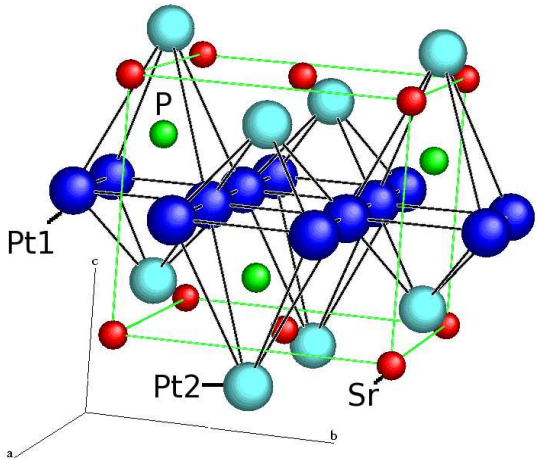}
}
\caption{Fig. 8. Crystal structure of SrPt$_3$P.  Pt$_6$P octahedra are
distinguished by Pt-Pt bonds.}
\label{str_aptp}
\end{figure}

{\em Electronic structure.}

At the Fermi level density of states is essentially determined by Pt1-5d 
states of Pt1 ions forming two-dimensional square lattice. 
(cf. Fig. 8). Also at the Fermi level there is a small admixture of
Pt2-5d and P-3p states.

The value of density of states at the Fermi level $N(E_F)$ in compounds with
Sr and La is equal to 4.69 states/eV/cell and 3.77 states/eV/cell
correspondingly. These values are comparable with those in HTSC pnictides with
relatively high  $T_c$ (cf. Ref. \cite{Kucinskii10}). Then the coefficient of
linear term in specific heat $\gamma_b$ is equal to 11 mJ/mole/K$^2$ and 
8.9 mJ/mole/K$^2$ for SrPt$_3$P and LaPt$_3$P respectively, which agree rather
well with experimental estimates for SrPt$_3$P giving
$\gamma^{exp}=$12.7 mJ/mole/K \cite{Takayama}. Strictly speaking
$\gamma^{exp}$ should be somehow larger, than those obtained in free-electron 
model, due to renormalization of the density of states by electron-phonon
interaction: $\gamma=(1+\lambda)\gamma_b$, where $\lambda$ is dimensionless
coupling constant. Comparing experimental and calculated values we can
estimate this constant as $\lambda$=0.15, corresponding to weak coupling, which
is insufficient to obtain the experimental value of $T_c$. Note, that according
to experimental estimates made in Ref. \cite{Takayama}, the so called Wilson
ratio $R_W\sim 1$, which supports the absence of strong correlation effects
in SrPt$_3$P.

In Fig. 9 we show band dispersions for  SrPt$_3$P (black lines) and LaPt$_3$P
(gray lines) in the vicinity of the Fermi level, obtained in LDA calculations.
These dispersions strongly differ from dispersions iné SrPt$_2$As$_2$ 
(cf. Fig. 6) \cite{srpt2as2}, and from those in HTSC pnictides and chalcogenides
(cf. Fig. 1) \cite{PvsC}. First of all, both APt$_3$P systems are
essentially three-dimensional, as can be seen from the presence of dispersion
along $\Gamma$-$Z$ direction. From chemical composition point of view (neglecting
the relaxation of lattice) the system LaPt$_3$P looks like electron doped
SrPt$_3$P system.

\begin{figure*}
\center{
\includegraphics[clip=true,width=0.6\textwidth]{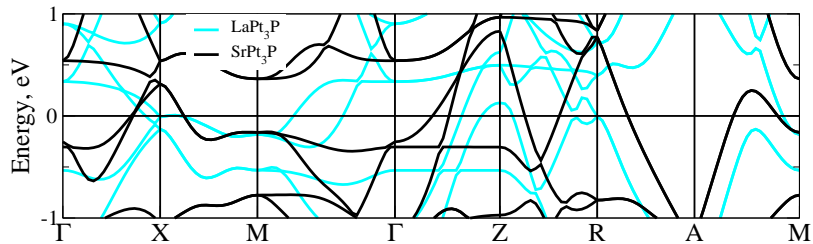}
}
\caption{Fig. 9.  LDA calculated electronic dispersions for SrPt$_3$P close to
the Fermi level (black lines) and LaPt$_3$P (gray lines).
Fermi level is at zero energy.
}
\label{band_aptp}
\end{figure*}

Fig. 10 shows Fermi surfaces for SrPt$_3$P (left) and LaPt$_3$P (right)
obtained by us via LDA calculations. In general, the form of Fermi surface of
APt$_3$P systems sharply differs from those in iron pnictides and chalcogenides
\cite{Nekr2,Nekr_kfese}. In particular, these Fermi surfaces are essentially
of three-dimensional nature and do not have well-defined cylinders.

\begin{figure}
\center{
\includegraphics[clip=true,width=0.2\textwidth]{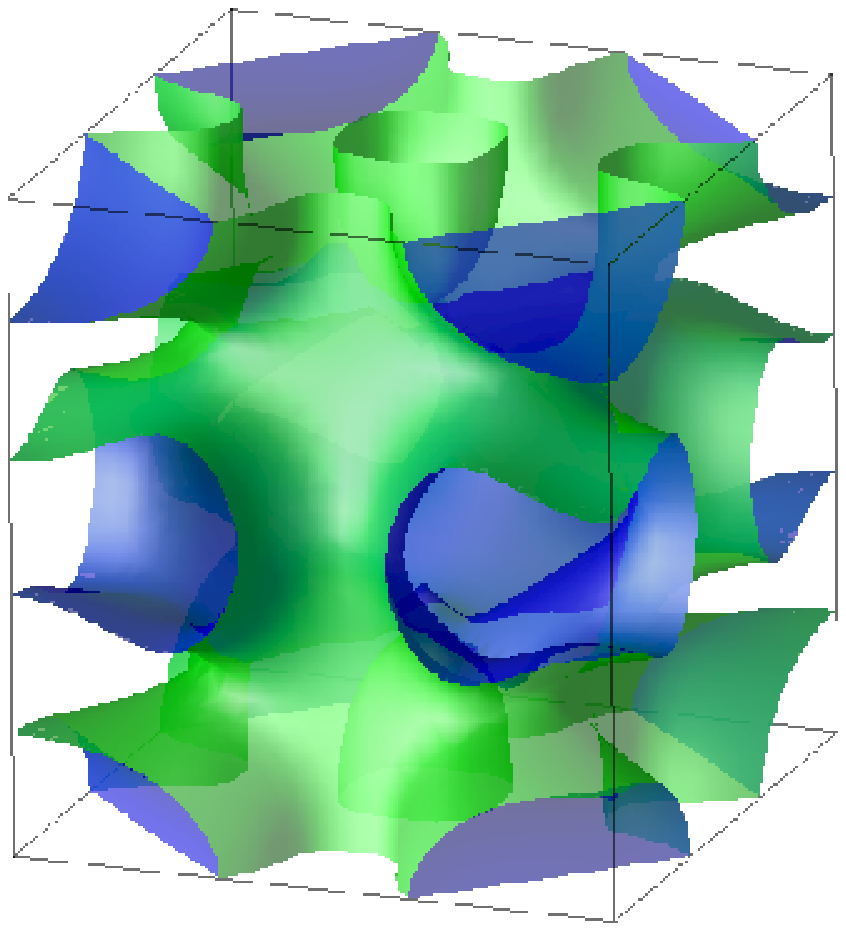}
\includegraphics[clip=true,width=0.2\textwidth]{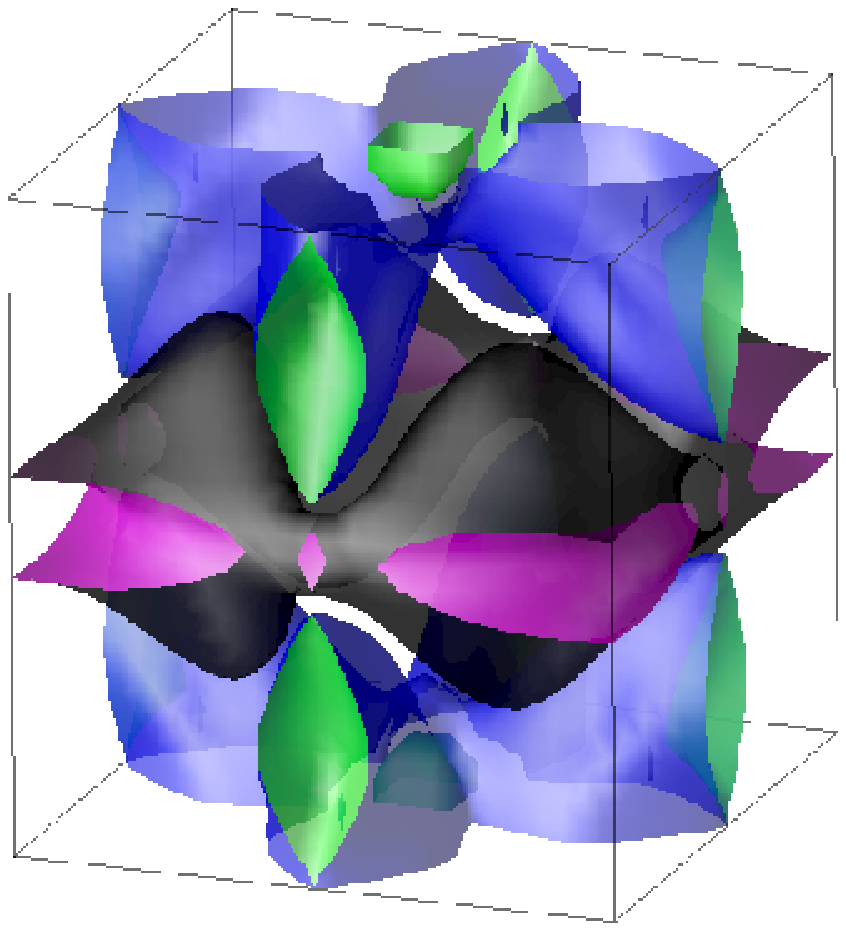}
}
\caption{Fig. 10. Fermi surfaces for SrPt$_3$P (left) and LaPt$_3$ (right) 
obtained from LDA calculations.
}
\label{fs1_aptp}
\end{figure}

From above discussion on electronic structure we can conclude, that APt$_3$P 
compounds represent a new class of systems with multiple-band superconductivity
as it was assumed in Ref. \cite{Takayama} from the measurements of Hall
coefficient. SrPt$_3$P is essentially a two-band superconductor, while
LaPt$_3$ possesses rather complicated band structure in the vicinity of the
Fermi level. Fermi surfaces of both systems are characterized by large number
of sheets with large number of ``pockets'' in the first Brillouin zone, with
topology only slightly changing under doping.

In this case, from general symmetry analysis \cite{VG,SU} it is again cleat,
that in case of singlet pairs, in principle, it is possible to have either
the usual isotropic or anisotropic $s$-wave pairing or several types of
$d$-wave pairing. Most probable is certainly the case of the usual
$s$-wave pairing, as was assumed in Ref. \cite{Takayama}. Additional
complications arise due to multiple-band nature of APt$_3$P compounds.
Three- dimensional structure of multisheet Fermi surface can lead to
complicated structure of superconducting gap with different values of
superconducting gaps on different sheets of the Fermi surface
(cf. discussion if Refs. \cite{Gork,KS09}). However, the simple BCS expression
for $T_c=1.14\omega_De^{-1/\lambda}$ allows to estimate the dimensionless
coupling constant $\lambda$ using the experimental values of $T_c$ and Debye
frequency $\omega_D$=190K \cite{Takayama}, which gives
$\lambda$=0.31 for $T_c$=8.4K. Lowering the value of $\lambda$
proportionally to the value of the density of states at the Fermi level,
which changes from 4.69 states/eV/cell in SrPt$_3$P to 3.77 states/eV/cell in 
LaPt$_3$P, we obtain $T_c$=4K for LaPt$_3$P, in reasonable agreement with
experimental value of 1.5K.

Simplest BCS expression for $T_c$ gives only a rough estimate and it is better
to use McMillan formula \cite{MM}. Rather close estimate can be also obtained 
from Allen-Dynes expression \cite{AD}, which is the best interpolation 
formula for $T_c$ for strong coupling superconductors. Let us choose first an
``optimistic'' value of Coulomb pseudopotential $\mu^*=0$. Then McMillan's
expression gives $\lambda$=0.61 for SrPt$_3$P and, in turn, the value of 
$T_c=5.6$K for LaPt$_3$P. Assuming more typical value of  $\mu^*=0.1$ 
we obtain  $\lambda$=0.85  for SrPt$_3$P, so that for LaPt$_3$P we getì 
$T_c$=5.4K.

In summary, we can conclude, that the value of $T_c$ in SrPt$_3$P and LaPt$_3$P
correlates rather well with the value of the total density of states at the
Fermi level, similarly to the case of HTSC pnictides and chalcogenides
\cite{PvsC,Kucinskii10}. At the same time, these estimates of $T_c$ 
correspond to weak or intermediate coupling superconductivity in APt$_3$P, 
which does not explain (cf. Ref. \cite{AD}) unusually high value of
$2\Delta/T_c$ ratio obtained from specific heat measurements in 
Ref. \cite{Takayama}, hinting to the necessity of additional experimental 
verification of the value of $2\Delta/T_c$ in these systems.

\subsection{BaFe$_2$Se$_3$ system}

BaFe$_2$Se$_3$ (Ba123) system \cite{Ba123_1} was synthesized as a possible
superconductor, analogous to (K,Cs)Fe$_2$Se$_2$ (with $T_c\sim$11K according to
preliminary data), though in a later work \cite{Ba123_2} superconductivity has
not been observed up to 1.8K. In both works neutron diffraction was used to
discover antiferromagnetic ``spin ladders'' with Neel temperature of the
order of 250Ê, though the magnetic structure was not determined unambiguously. 
In this section we describe both electronic and magnetic structures of Ba123, 
obtained from LDA and LSDA calculations in Ref. \cite{Ba123}.

{\em Crystal structure.}

We start as usual from description of crystal structure of Ba123.
Ba123 system possesses orthorhombic symmetry group {\em Pnma} \cite{Ba123_1}.
Basic structure elements in this compound are the so called
``two-leg'' ladders oriented along $b$-axis. These ladders are formed by iron
ions, surrounded by tetrahedra formed by Se ions In orthogonal direction to.
$b$-axis these ladders are placed in chessboard order (cf. Ref. \cite{Ba123}).
Obviously, this crystal structure is sharply different from crystal structures
of iron based HTSC systems (cf. Section 2 and Refs. \cite{PvsC,Nekr2,Nekr_kfese}),
which belong to primitive or body-centered tetragonal symmetry group.

{\em Electronic structure.}

In Fig. 11 we show band dispersions (right) and densities of states
(left) for Ba123 system Similarly to  Ba122 systems \cite{Nekr2} and 
AFe$_2$Se$_2$ \cite{Nekr_kfese} electronic states at the Fermi level are 
formed mainly by Fe-3d orbitals. Se-4p orbitals form bands placed -2 eV lower
in energy. Hybridization between Fe-3d and Se-4p states is relatively 
moderate.

\begin{figure}
\center{
\includegraphics[clip=true,width=0.5\textwidth]{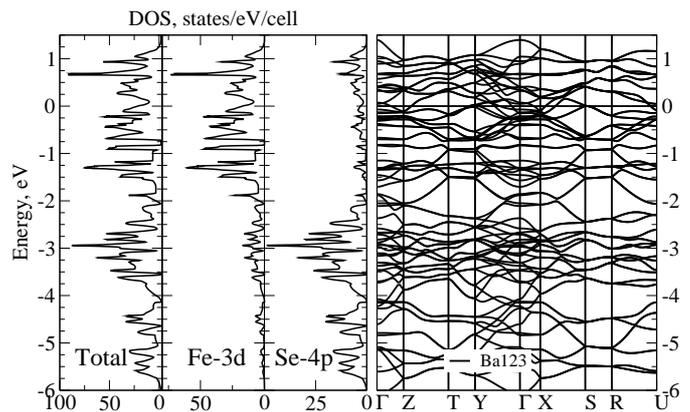}
}
\label{band_bafese}
\caption{Fig. 11. Band dispersions and densities of states for Ba123
compound, obtained in LDA calculations. Fermi level -- zero energy.}
\end{figure}

Electronic bands of Ba123 system in close vicinity of the Fermi level strongly
differ from those in iron pnictides and chalcogenides.
\cite{Nekr2,Nekr_kfese}. Around $\Gamma$-point we have two electronic pockets,
while at the Brillouin zone edges (Y-point) there are three hole pockets
(cf. Fig. 12). Close to the Fermi level we observe several Van-Hove
singularities, making this system similar to AFe$_2$Se$_2$ (cf. Section 2) 
\cite{Nekr_kfese} and allowing changing Fermi surface topology by doping.

Fermi surface following from LDA calculations is shown in Fig. 12.
The general form of this Fermi surface is completely different from the case
of iron pnictides and chalcogenides \cite{Nekr2,Nekr_kfese}. In Ba123 it is 
essentially three-dimensional and without explicit cylinders

\begin{figure}
\center{
\includegraphics[clip=true,width=0.45\textwidth]{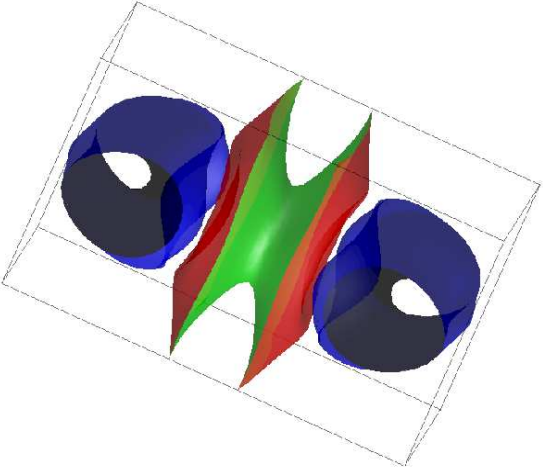}
}
\caption{Fig. 12. Fermi surface of Ba123 obtained in LDA calculations.}
\label{fs_bafese}
\end{figure}

{\em Magnetic structure.}

Neutron scattering experiments \cite{Ba123_1,Ba123_2} allowed to determine
Neel temperature $T\ll T_N^{exp}\sim$250K in Ba123, as well as two possible
spin configurations, corresponding to irreducible representations
$\tau_1$ and $\tau_2$ of space symmetry group {\em Pnma}, which produce
practically identical diffraction patterns. In Ref. \cite{Ba123} the problem
of the type of magnetic structure was solved by direct calculation of
Neel temperatures for different spin configurations in mean-field approximation
for Heisenberg model. Parameters of Heisenberg model were calculated within
LSDA \cite{leip}. Our calculations demonstrated, that spin ladder configuration
$\tau_1$ (``plaquettes'') has Neel temperature $T_N(\tau_1)$=217K, while for
$\tau_2$ (``zigzags'') -- $T_N(\tau_2)$=186K, which makes $\tau_2$ configuration
more favorable energetically, in accordance wit experimental 
work~\cite{Ba123_2} and LSDA calculations of total energy~\cite{Saparov}.

Thus, despite similar to iron based HTSC chemical composition, Ba123 system
possesses essentially different crystal and electronic structure, while from
the point of view of magnetism it is a ``spin ladder''. The question of
superconductivity in this system remains open.

\subsection{APd$_2$As$_2$ system}

This section is devoted to electronic structure 
of (Sr,Ca)Pd$_2$As$_2$ and  BaPd$_2$As$_2$ systems  \cite{srpd2as2}, 
where superconductivity was discovered with T$_c$ 0.92K and 1.27K respectively~\cite{Anand}.
We also compare these systems with isovalent system
(Sr,Ba)Ni$_2$As$_2$ \cite{Subedi_Ni,Zhou_Ni,Shein_Ni}. 

{\em Crystal structure.}

Crystal structure of SrPd$_2$As$_2$ and CaPd$_2$As$_2$ belongs to tetragonal
space symmetry group $I/4mmm$, similarly to BaFe$_2$As$_2$ system \cite{rott}. 
We show this structure in the left part of Fig.13. We can see, that it is
analogous to Ba122 system \cite{Nekr2}.
\begin{figure}[ht]
\begin{center}
\includegraphics[clip=true,width=0.22\textwidth]{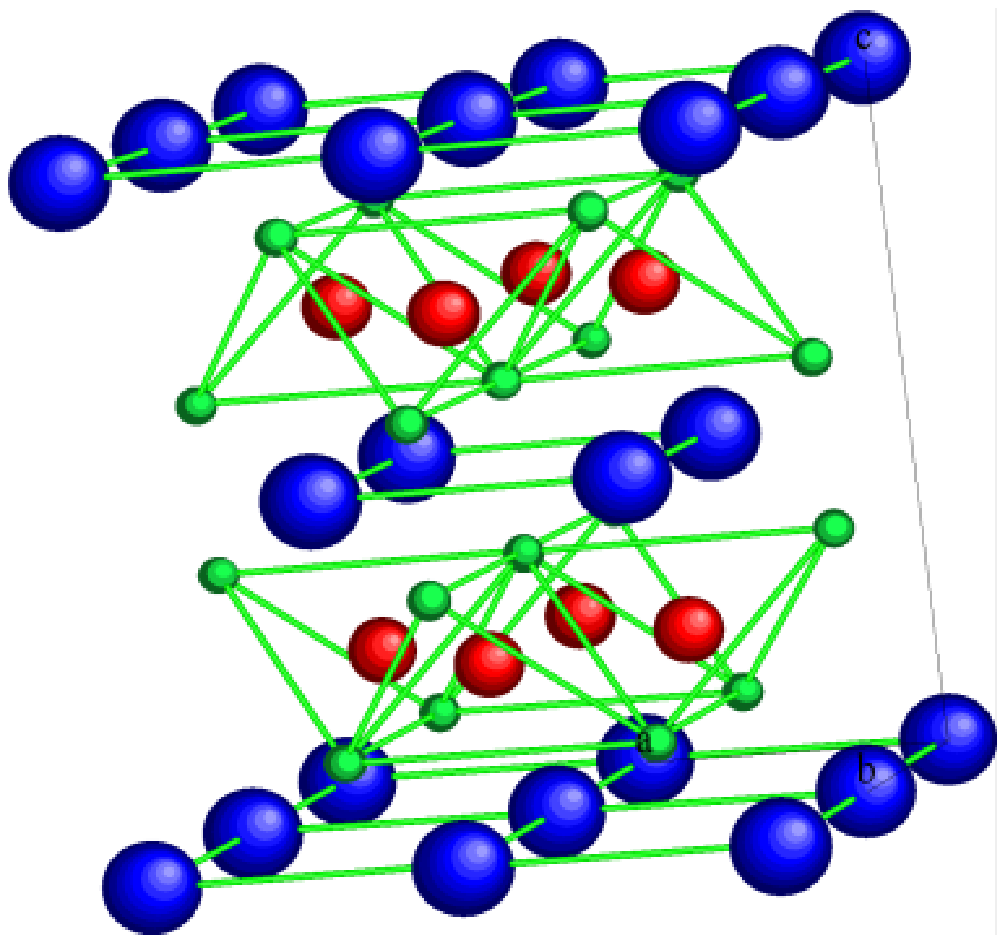}
\includegraphics[clip=true,width=0.2\textwidth]{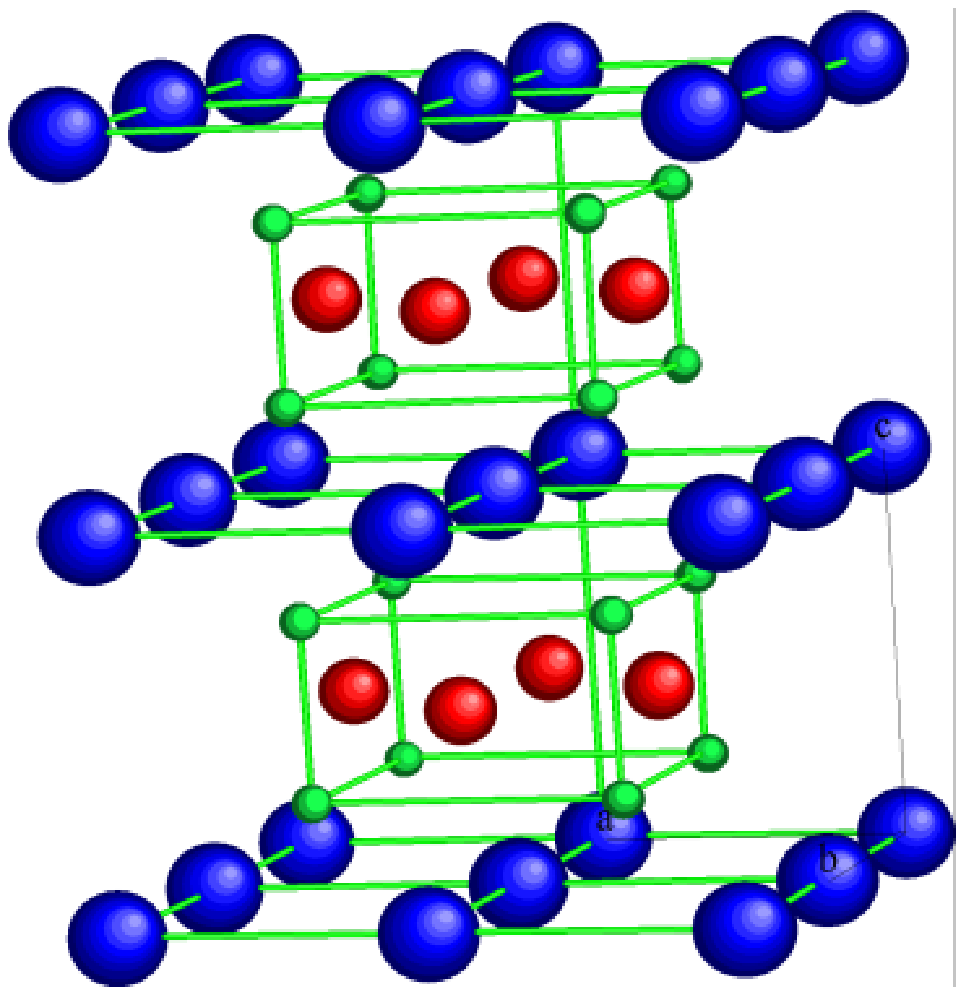}
\label{fig1}
\end{center}
\caption{Fig. 13. Crystal structure of (Sr,Ca)Pd$_2$As$_2$ (left) and 
BaPd$_2$As$_2$ (right).
Blue spheres -- ions of Sr and  Ba , green -- As and red  -- Pd.} 
\end{figure}

Though chemical formula of BaPd$_2$As$_2$ is similar to that of
Ba122 pnictide, it has completely different crystal structure \cite{Anand}.
Space symmetry group of  BaPd$_2$As$_2$ ñèñòåìû is $P/4mmm$ and its structure
is shown in right part of Fig. 13. This crystal structure, similarly to iron
pnictides and chalcogenides, is a layered one, but Pd atoms do not possess
surrounding tetrahedra of As atoms.

{\em Electronic structure.}

LDA calculated electronic dispersions along high-symmetry directions
for SrPd$_2$As$_2$ and BaPd$_2$As$_2$, as well as densities of states, are 
shown in upper and lower parts of Fig. 14. From densities of states it can be 
seen, that in SrPd$_2$As$_2$ system, the main part of spectral weight is
formed by Pd-4d and As-4p states.  Pd-4d states belong to energy interval
between -4 è -0.5 ýÂ (cf. upper part of Fig. 14), while As-4p states are in
the energy interval (-6;-4) eV. As compared with ñ Ba122 Pd-4d states
are more extended in energy, than Fe-3d states. Also in SrPd$_2$As$_2$ 
we have considerable hybridization between Pd-4d and As-4p states. The value of
the density of states of SrNi$_2$As$_2$ at the Fermi level 
$N(E_F)$=1.93 states/eV/cell. which is twice lower, than in
Ba122 system, due to the larger number of electrons in Pd, as compared to Fe, 
leading to the shift of the Fermi level to the region of lower density of
states.

\begin{figure}
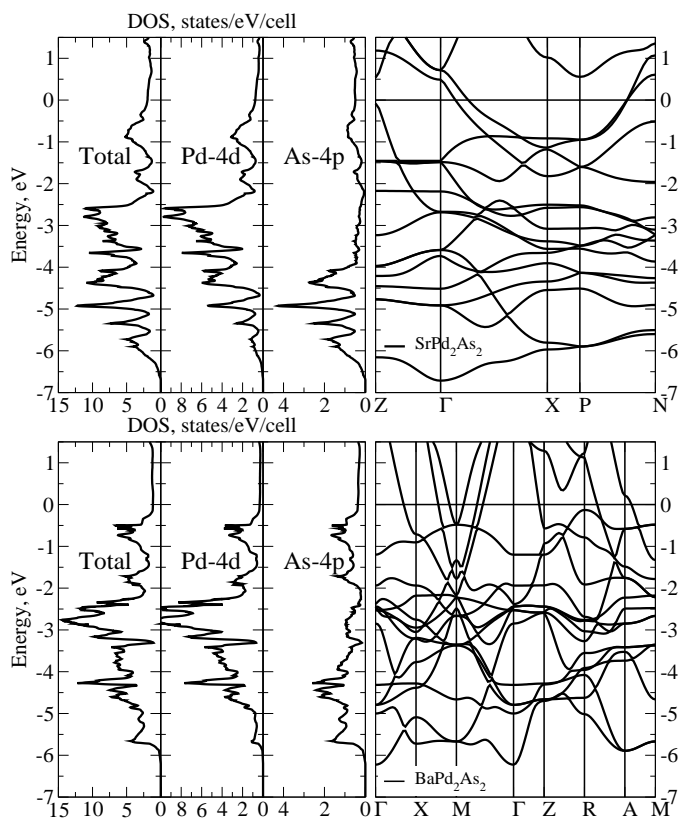

\includegraphics[clip=true,width=0.5\textwidth]{Sr_DOS_bands.eps}
\includegraphics[clip=true,width=0.5\textwidth]{Ba_DOS_bands.eps}
\caption{Fig. 14. Electronic dispersions and densities of states for
SrPd$_2$As$_2$ (above) and BaPd$_2$As$_2$ (below) calculated in â LDA.
Fermi level  -- zero energy.} 
\end{figure}

In the lower part of Fig. 14 we show electronic dispersions and densities of 
states obtained for BaPd$_2$As$_2$ in LDA calculations. As the crystal structure 
of BaPd$_2$As$_2$ is completely different from that of SrNi$_2$As$_2$, it is of
no surprise, that bands are also different. Note, that in Sr system, as well as
in Ba system, the Fermi level is crossed by multiple bands, without any clear
contributions from different valence shells. From LDA calculations we have
found the value of total density of states at the Fermi level of
BaPd$_2$As$_2$ to be 2.29 states/eV/cell.

In Fig. 15 we show Fermi surfaces for SrPd$_2$As$_2$ (left) and
BaPd$_2$As$_2$ (right), obtained from our band structure calculations.
Fermi surface for  Sr system is essentially three-dimensional and complex, in
contrast to Ba122 system \cite{Nekr2}, and consists of three sheets.
Fermi surface of BaPd$_2$As$_2$, as compared with that of SrPd$_2$As$_2$, is
rather simple, though also three-dimensional, with hole-like sheet around
$\Gamma$-point and electron=like sheets at the corners of Brillouin zone.

\begin{figure}
\begin{center}
\includegraphics[clip=true,width=0.15\textwidth]{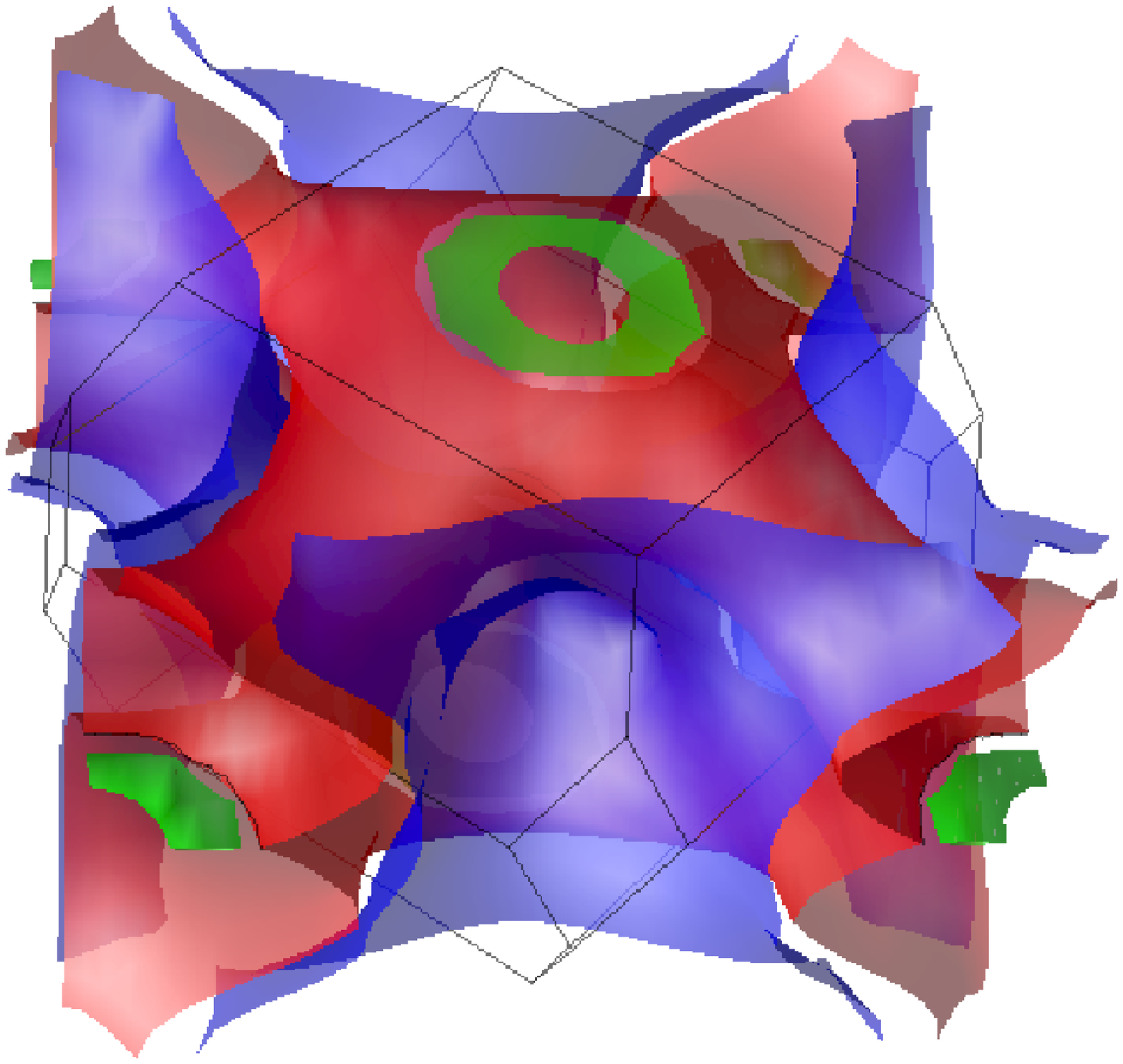}
\includegraphics[clip=true,width=0.2\textwidth]{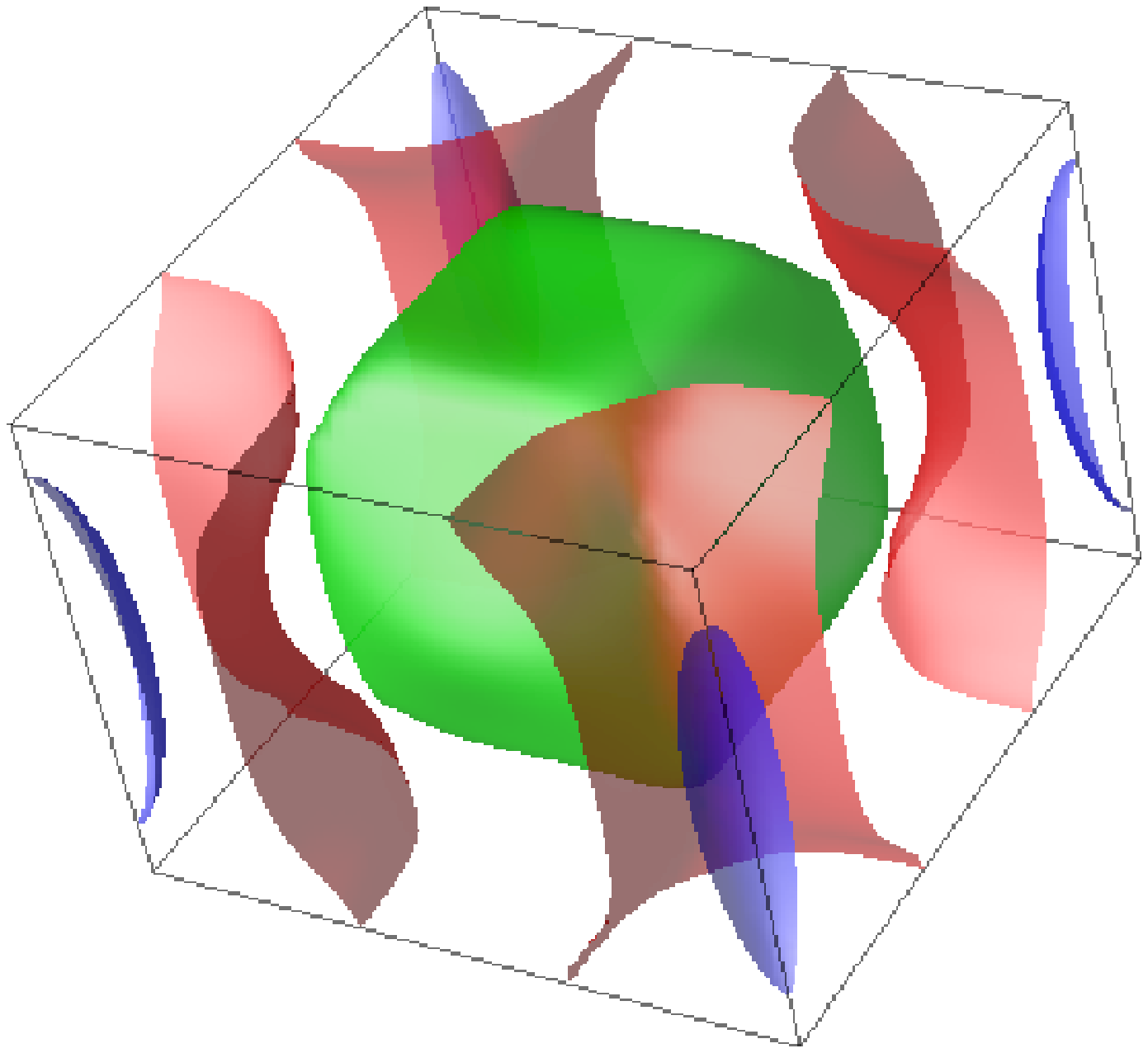}
\end{center}
\caption{Fig. 15. LDA calculated Fermi surfaces for (Sr,Ca)Pd$_2$As$_2$ (left) 
and BaPd$_2$As$_2$ (right).
} 
\end{figure}

\section{Conclusion}

In this small review we considered electronic structure of a number of new
superconductors, discovered during the experimental search of new candidate
systems for high-temperature superconductivity, which followed the
observation of HTSC in iron arsenides \cite{kamihara_08, UFN_90, Hoso_09, MazKor}.
All these systems are multiple-band superconductors with rather complex
topology of Fermi surfaces. From general theoretical considerations \cite{KS09} 
it is clear, that multiple-band structure of electronic spectrum facilitates
raising the temperature of superconducting phase transition $T_c$. 
However, the experimental picture is rather contradictory. 
In K$_{1-x}$Fe$_{2-y}$Se$_2$ system superconductivity is observed at high
enough temperatures, despite the absence of ``nesting'' of electron-like and
hole-like Fermi surfaces, as well as the absence of well-defined quasiparticles
close to the Fermi level (due to the strong role of electronic correlations in
this systems). At the same time, the majority of other systems, considered 
above, demonstrate superconductivity at relatively low temperatures, despite 
clearly multiple-band nature of electronic spectrum.

Because of this contradictory picture, the general question arises, whether
there exists (or does not exist) some electronic structure, which is somehow
``optimal'' from the point of view of observation of HTSC. At present we can not
give any definite answer to this question, though from our calculations it can
be seen, that electronic spectra of the majority of systems with low $T_c$,
considered above, have spectra sharply different from those in iron pnictide
and iron chalcogenide HTSC systems. Attempts to find the general answer on the
the posed question seem to be an interesting direction of further investigations.

This work is partially supported by RFBR grant 14-02-00065, as well as by the
grants of the programs of fundamental research of the Presidium of UB RAS
``Quantum macrophysics and nonlinear dynamics'' (projects No. 12-$\Pi$-2-1002, 
12-T-2-1001).

\end{document}